\documentclass[aps,pra,twocolumn,showpacs,amsmath,amsmath,amssymb,amsfonts,superscriptaddress]{revtex4-1}
\usepackage{amssymb}
\usepackage{graphicx}
\usepackage{dcolumn}
\usepackage{bm}
\usepackage{psfrag}
\usepackage{bbold}
\usepackage{float}
\usepackage{amsmath,braket}

\usepackage{xcolor}

\def\beq{\begin{equation}}
\def\eeq{\end{equation}}
\def\beqa{\begin{eqnarray}}
\def\eeqa{\end{eqnarray}}

\begin{document}

\author{Robert J. Bettles}
\affiliation{Joint Quantum Center (JQC) Durham-Newcastle, Department of Physics, Durham University, South Road, Durham DH1 3LE, United Kingdom}
\author{Ji\v{r}\'{i} Min\'{a}\v{r}}
\affiliation{School of Physics and Astronomy, The University of Nottingham, Nottingham, NG7 2RD, United Kingdom}
\affiliation{Centre for the Mathematics and Theoretical Physics of Quantum Non-equilibrium Systems, The University of Nottingham, Nottingham, NG7 2RD, UK}
\author{Charles S. Adams}
\affiliation{Joint Quantum Center (JQC) Durham-Newcastle, Department of Physics, Durham University, South Road, Durham DH1 3LE, United Kingdom}
\author{Igor Lesanovsky}
\affiliation{School of Physics and Astronomy, The University of Nottingham, Nottingham, NG7 2RD, United Kingdom}
\affiliation{Centre for the Mathematics and Theoretical Physics of Quantum Non-equilibrium Systems, The University of Nottingham, Nottingham, NG7 2RD, UK}
\author{Beatriz Olmos}
\email{beatriz.olmos-sanchez@nottingham.ac.uk}
\affiliation{School of Physics and Astronomy, The University of Nottingham, Nottingham, NG7 2RD, United Kingdom}
\affiliation{Centre for the Mathematics and Theoretical Physics of Quantum Non-equilibrium Systems, The University of Nottingham, Nottingham, NG7 2RD, UK}

\title{Topological properties of a dense atomic lattice gas}
\date{\today}
\begin{abstract}
We investigate the existence of topological phases in a dense two-dimensional atomic lattice gas. The coupling of the atoms to the radiation field gives rise to dissipation and a non-trivial coherent long-range exchange interaction whose form goes beyond a simple power-law. The far-field terms of the potential -- which are particularly relevant for atomic separations comparable to the atomic transition wavelength -- can give rise to energy spectra with one-sided divergences in the Brillouin zone. The long-ranged character of the interactions has another important consequence: it can break of the standard bulk-boundary relation in topological insulators. We show that topological properties such as the transport of an excitation along the edge of the lattice are robust with respect to the presence of lattice defects and dissipation. The latter is of particular relevance as dissipation and coherent interactions are inevitably connected in our setting.
\end{abstract}


\maketitle

\noindent\textit{Introduction.}
Recently, the pursuit of topological phases in quantum many-body systems has been the focus of intense research. The potential application of these topological states for robust quantum computation \cite{Kitaev03,Nayak08} is one of the driving forces for this increased interest. So-called topological insulators are usually characterized by bulk bands separated by a gap and the presence of gapless edge states whose properties are topologically protected against local perturbations such as external disorder or noise \cite{Hasan10,Yao13}. Paradigmatic examples of these include the integer and fractional quantum Hall effects, which were initially realized on two-dimensional electron gases subject to strong magnetic fields \cite{Klitzing80,Laughlin81,Tsui82,Laughlin83}. Since their discovery, several lattice models that do not require external magnetic fields have been proposed and some realized experimentally \cite{Haldane88,Ruostekoski02,Kane05,Novoselov07,Tsukazaki07,Konig07,Raghu08,Neupert11,Wang11,Tang11,Sun11,Wang12,Dauphin12,Cooper13,Barkeshli15,Andrijauskas15}.

Due to the high degree of experimental control that is achievable nowadays, cold atoms and molecules have been proposed as platforms for the exploration of novel topological phases of quantum matter \cite{Goldman16}. In particular, many-body systems that display long-range interactions \cite{Zhang14,Vodola14,Gong16,Behrmann16} -- such as polar molecules \cite{Peter12,Yao12,Manmana13,Yao13_1,Peter15}, atoms with large magnetic dipoles \cite{Yao15}, and Rydberg atoms \cite{Maghrebi15} -- have been shown to feature topologically non-trivial flat bands and fractional quantum Hall states.

\begin{figure}
\centering
 \includegraphics[width=1\columnwidth]{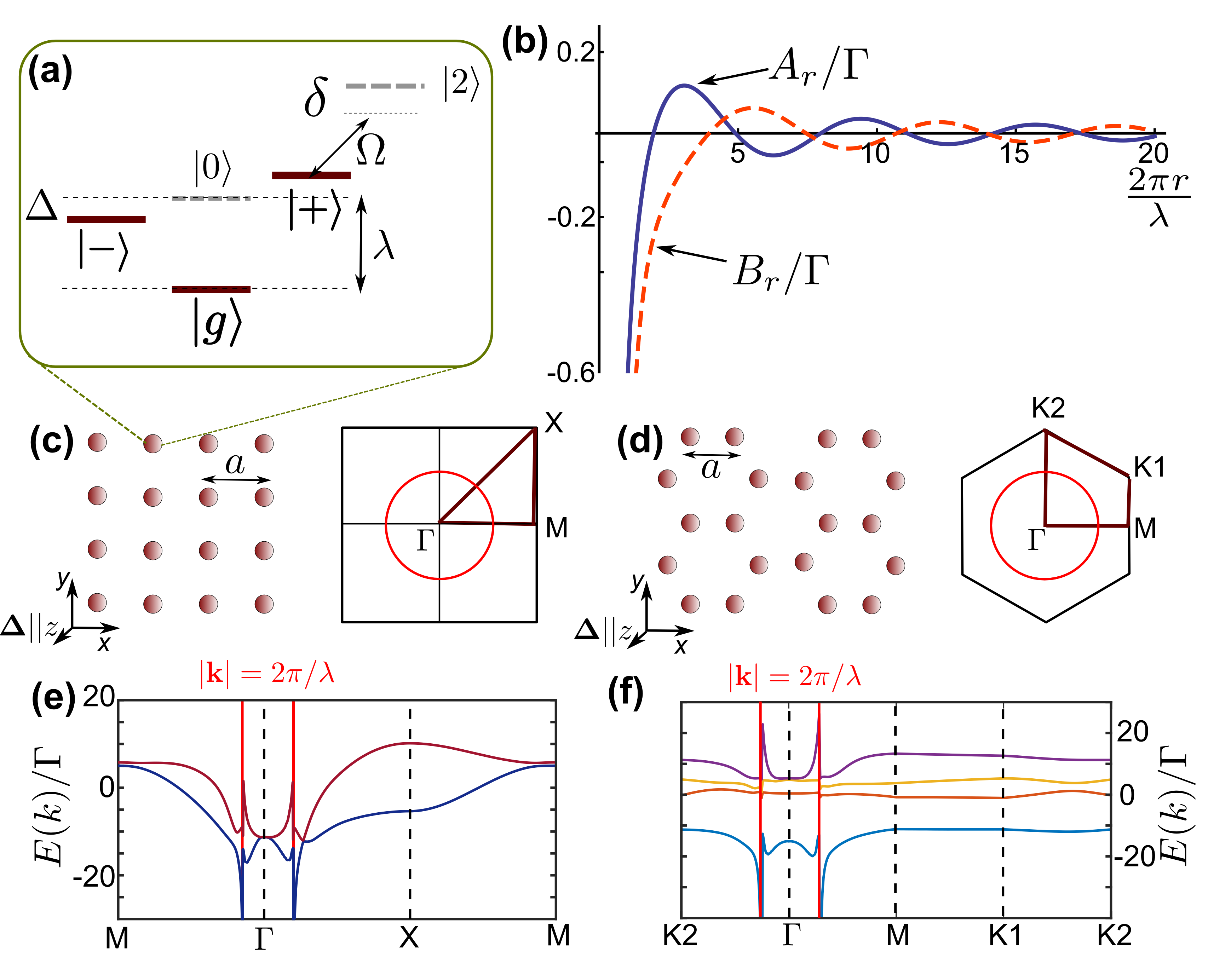}\\
 \caption{\textbf{(a):} Internal atomic level structure. Time reversal symmetry is broken via a magnetic field of strength $\Delta$ perpendicular to the lattice and the off-resonant coupling of $\left|+\right>$ to an auxiliary state $\ket{2}$ with $\varepsilon\approx\Omega/2\delta\ll 1$.
 \textbf{(b):} Hamiltonian coefficients $A_r$ [Eq. (\ref{eq:A}), solid line] and $B_r$ [Eq. (\ref{eq:B}), dashed line] vs. $r/\lambda$. \textbf{(c)} and {\bf (d):} Real space configurations and first Brillouin zone for SL and HL, respectively. The red circle represents the points of divergence and discontinuity of the spectrum.
  \textbf{(e)} and {\bf(f):} Dispersion relations for the SL and the HL, respectively, which exhibit the existence of a one-sided divergence at $|\mathbf{k}|=2\pi/\lambda$.}\label{fig:scheme}
\end{figure}

In this paper, we explore the topological properties of a two-dimensional lattice of atoms where long-range interactions arise intrinsically via coherent light scattering between internal atomic states \cite{Bettles15}. We consider the full interaction -- going thus beyond the usually employed near-field approximation -- which is of relevance for atomic separations comparable to the atomic transition wavelength. This scenario is currently studied in various contexts, e.g., the exploration of collective light scattering and super- and sub-radiant decay \cite{Svidzinsky10,Keaveney12,Bienaime13,Pellegrino14,Kwong14,Bettles15,Bromley16,Jenkins16}. We study two different lattice geometries (a square and a honeycomb lattice) and find that this simple system can support topologically non-trivial phases. The often neglected far-field terms of the interactions lead to one-sided divergences in the single-particle spectrum. We explore the consequences of the long-range character of the interaction on the relation between the topological properties of the bulk and the boundary of a finite size system. Remarkably, we find that the standard bulk-boundary relation \cite{Hatsugai93,hatsugai93-2}, which is well established for short-range interactions, does not generally hold. Furthermore, we find that neither the presence of lattice defects nor dissipation destroy the topological properties of the system. The latter is of particular importance, as the considered system is inevitably open: both dissipation and coherent exchange interactions originate from the coupling of the atoms to the radiation field.

\noindent\textit{The system.}
We consider a two-dimensional optical lattice with $N$ sites lying in the $xy$ plane (lattice spacing $a$), where each site is occupied by a single atom. The $j$-th atom is located at position ${\mathbf r}_j$ with internal levels $\ket{g}_j$ (the ground state) and $\ket{-}_j, \ket{0}_j, \ket{+}_j$ (the excited states) corresponding to the $J=0$ and $J=1$ total angular momentum manifolds, respectively [see Fig. \ref{fig:scheme}(a)]. This level structure is naturally available in a variety of systems such as alkaline-earth-metal atoms \cite{Gorshkov09,Gorshkov10,Olmos13}, dysprosium atoms \cite{Yao15}, polar molecules \cite{Peter15,Gorshkov13,Syzranov14} or Rydberg systems \cite{Maghrebi15}.

The coupling of the atoms to the quantized multimode radiation field results in an effective long-range exchange interaction and collective dissipation \cite{Lehmberg70,Agarwal70,James93}. Within the dipole and Born-Markov approximations, the dynamics of the atomic system is described by the master equation $\dot{\rho}=-\frac{i}{\hbar}\left[H,\rho\right]+{\cal D}(\rho)$, with
\begin{eqnarray}
	H &=& \hbar\sum_{j\neq l}\mathbf{d}^\dag_{j}\cdot\overline{V}_{jl}\cdot\mathbf{d}_{l} \label{eq:H} \\
	{\cal D}(\rho) &=& \sum_{jl}\mathbf{d}_j\cdot\overline{\Gamma}_{jl} \cdot\rho\mathbf{d}_l^\dag-\frac{1}{2}\left\{\mathbf{d}_j^\dag\cdot\overline{\Gamma}_{jl} \cdot\mathbf{d}_l,\rho\right\}. \label{eq:diss}
\end{eqnarray}
Here, $\mathbf{d}_j=\left(\left|g\right>_j\!\left<+\right|,\left|g\right>_j\!\left<-\right|\right)^T$ and
\begin{eqnarray}
  \overline{V}_{jl}&=&\left(\begin{array}{cc}
    A_{jl} & B_{jl}e^{-2i\phi_{jl}}\\
    B_{jl}e^{2i\phi_{jl}} & A_{jl}
  \end{array}\right) \label{eq:W} \\
  \overline{\Gamma}_{jl}&=&\left(\begin{array}{cc}
    A'_{jl} & B'_{jl}e^{-2i\phi_{jl}}\\
    B'_{jl}e^{2i\phi_{jl}} & A'_{jl}
  \end{array}\right). \label{eq:Gamma}
\end{eqnarray}
Note, that we have used here the fact that the dynamics of the $\ket{0}$ level and the $\{\ket{-},\ket{+}\}$ manifold are decoupled, see Appendix A for details. The coefficients in Eqs. (\ref{eq:W}) and (\ref{eq:Gamma}) read
\begin{eqnarray}
  A_{jl}&=&\frac{3\Gamma}{8}\left[-\frac{\cos{\kappa_{jl}}}{\kappa_{jl}}-\frac{\sin{\kappa_{jl}}}{\kappa_{jl}^2} -\frac{\cos{\kappa_{jl}}}{\kappa_{jl}^3}\right]\label{eq:A}\\
  B_{jl}&=&\frac{3\Gamma}{8}\left[\frac{\cos{\kappa_{jl}}}{\kappa_{jl}}-3\left(\frac{\sin{\kappa_{jl}}}{\kappa_{jl}^2} +\frac{\cos{\kappa_{jl}}}{\kappa_{jl}^3}\right)\right]\label{eq:B}\\
  A'_{jl}&=&\frac{3\Gamma}{4}\left[\frac{\sin{\kappa_{jl}}}{\kappa_{jl}}-\frac{\cos{\kappa_{jl}}}{\kappa_{jl}^2} +\frac{\sin{\kappa_{jl}}}{\kappa_{jl}^3}\right]\label{eq:Ap}\\
  B'_{jl}&=&\frac{3\Gamma}{8}\left[-\frac{\sin{\kappa_{jl}}}{\kappa_{jl}}-3\left(\frac{\cos{\kappa_{jl}}}{\kappa_{jl}^2} -\frac{\sin{\kappa_{jl}}}{\kappa_{jl}^3}\right)\right],\label{eq:Bp}
\end{eqnarray}
where $\Gamma$ is the single atom decay rate and $\kappa_{jl}\equiv 2\pi r_{jl}/\lambda$, with $\lambda$ being the wavelength of the transition from the ground to the excited state manifold [see Fig. \ref{fig:scheme}(a)], $r_{jl}=|\mathbf{r}_{j}-\mathbf{r}_{l}|$, and $\phi_{jl}={\rm arg}(\mathbf{r}_{j}-\mathbf{r}_{l})$ is the polar angle between the $j$-th and $l$-th atom.

The many-body Hamiltonian (\ref{eq:H}) conserves the number of excitations in the system and describes their exchange among the atoms. The coefficients in this Hamiltonian [Eqs. (\ref{eq:A}) and (\ref{eq:B})] decay as $1/r^3$ for short distances ($r\leq\lambda$) and as $1/r$ for $r\gg\lambda$ [see Fig. \ref{fig:scheme}(b)]. We will study the topological properties of the system in an intermediate regime, where the full potential needs to be considered. In particular, in the remainder of the paper we will fix the ratio $a/\lambda=0.1$ \cite{Olmos13,Pellegrino14} (note that the effects we show are not constrained to this specific value) and systematically investigate two specific lattice geometries, the square lattice (SL) and the honeycomb lattice (HL) [see Figs. \ref{fig:scheme}(c) and (d), respectively]. We will proceed by analyzing first the band structure of the Hamiltonian (\ref{eq:H}) and then studying the effect of the dissipation (\ref{eq:diss}) on the edge transport in the topologically non-trivial phases.

In its current formulation, the Hamiltonian (\ref{eq:H}) is symmetric under time reversal and hence is topologically trivial. In order to possibly find topologically non-trivial phases we consider two ways of breaking the time reversal symmetry. First, we lift the degeneracy of the states $\left|\pm\right>$, which can be achieved by means of a magnetic field such that the two states' energies are shifted by $\pm\Delta$ [see Fig. \ref{fig:scheme}(a)]. Secondly, we couple off-resonantly the state $\left|+\right>$ via a microwave field to an auxiliary hyperfine state, $\left|2\right>$ \cite{Peter15}. Up to second order in $\varepsilon\approx\Omega/2\delta\ll1$, where $\Omega$ and $\delta$ are the Rabi frequency and the detuning of the microwave field, respectively, the $\left|+\right>$ state gets dressed and the potential (\ref{eq:W}) is modified as
\begin{equation}\label{eq:W-mod}
  \overline{V}_{jl}\!=\!\left(\!\!\begin{array}{cc}
    \left(A_{jl}+\Delta\delta_{jl}\right)\!\!\left(1-\varepsilon^2\right) & \!\!B_{jl}e^{-2i\phi_{jl}}\!\!\left(1-\frac{\varepsilon^2}{2}\right)\\
    B_{jl}e^{2i\phi_{jl}}\!\!\left(1-\frac{\varepsilon^2}{2}\right) & A_{jl}-\Delta\delta_{jl}
  \end{array}\!\!\right)\!,
\end{equation}
where $\delta_{jl}$ is the Kronecker delta symbol.

\noindent\textit{Band structure and divergences.} To obtain the band structure we express (\ref{eq:H}) in the reciprocal space as
\begin{equation}
	H=\hbar \sum_{\mathbf{k}} \psi_\mathbf{k}^\dag\cdot\overline{V}_\mathbf{k}\cdot\psi_{\mathbf{k}},
	\label{eq:Hk}
\end{equation}
where $\psi_\mathbf{k}=\frac{1}{\sqrt{N}}\sum_{l}\mathbf{d}_le^{i\mathbf{k}\cdot\mathbf{r}_{l}}$ and
\begin{equation}\label{eq:Wk}
\overline{V}_\mathbf{k}=\sum_{\mathbf{r}\neq0}e^{i\mathbf{k}\cdot\mathbf{r}}\overline{V}_\mathbf{r}
\end{equation}
with $\mathbf{r}$ being the relative distance between any pair of atoms. Note that the potential (\ref{eq:W-mod}) (due to the terms that decay as $1/r$) makes the sum in Eq. (\ref{eq:Wk})) not convergent. Potentials with similar features have been shown to lead to interesting physical effects, such as supersonic spreading of the correlations \cite{Raghu08,Neupert11,Wang11,Tang11,Sun11,Wang12,Dauphin12,Cooper13,Storch15}. Such potentials are also encountered in self-gravitating systems \cite{Campa14} or in electrons in solids subject to the Coulomb force \cite{Kantorovich04}.

To evaluate the sum (\ref{eq:Wk}) we employ the Ewald summation technique \cite{Bonsall77,Kantorovich04}, see Appendix B for details. As a result, we find that the potential (\ref{eq:Wk}) features one sided divergences in the spectrum occurring for wavevectors lying on a circle of radius $\kappa=2\pi/\lambda$ in reciprocal space. More specifically, the potential diverges as $\lim_{k \to \kappa^+} \braket{V_{\mathbf k}} = -\infty$, while $\lim_{k \to \kappa^-} \braket{V_{\mathbf k}}$ remains finite (see Appendix B). This situation is depicted in Fig. {\ref{fig:scheme}}(e) (SL) and Fig. {\ref{fig:scheme}}(f) (HL). Note that in this work we only study cases where the divergence falls inside the Brillouin zone, $a/\lambda<1/2$ for SL and $a/\lambda<1/3$ for HL.

\noindent\textit{Chern numbers and edge states.}
In the language of differential geometry, the topological properties are studied in terms of differentiable fibre bundles assuming a differentiable Hamiltonian map $H: T^2 \rightarrow \mathcal{M}$, mapping the Brillouin zone (represented as a two-dimensional torus $T^2$) to some target space $\mathcal{M}$ \cite{Nakahara03}. Here the topology is characterized by the Chern number defined as
\begin{equation}
	C=\frac{1}{2\pi i}\int_{T^2}d\mathbf{k}F_{xy}(\mathbf{k}),
	\label{eq:C}
\end{equation}
where $F_{xy}(\mathbf{k})=\partial_xA_y(\mathbf{k})-\partial_yA_x(\mathbf{k})$, $A_\mu(\mathbf{k})=\langle n(\mathbf{k})|\partial_\mu|n(\mathbf{k})\rangle$ is the Berry connection, $\ket{n(\mathbf{k})}$ is an eigenstate of the Hamiltonian (\ref{eq:Hk}), $\partial_\mu=\partial/\partial k_\mu,\; \mu=\{x,y\}$, and ${\mathbf k} \in T^2$.

In the present case the assumption of differentiability is not satisfied due to the discontinuity of the potential (\ref{eq:Wk}). However, formally it is still possible to evaluate the Chern number using the algorithm of Ref. \cite{Fukui05}, where special care is taken to avoid the points in the Brillouin zone where $\overline{V}_{\mathbf k}$ diverges. This corresponds to evaluating (\ref{eq:C}) with an effectively bounded Hamiltonian, which in turn yields integer values of $C$. Note also that the divergence appears as a consequence of considering the potential in the thermodynamic limit, a situation that is ill-defined as the finite propagation time of the radiation modes mediating the exchange of photons in the system is neglected \cite{Lehmberg70,Milonni74}.

\begin{figure}
\centering
 \includegraphics[width=1\columnwidth]{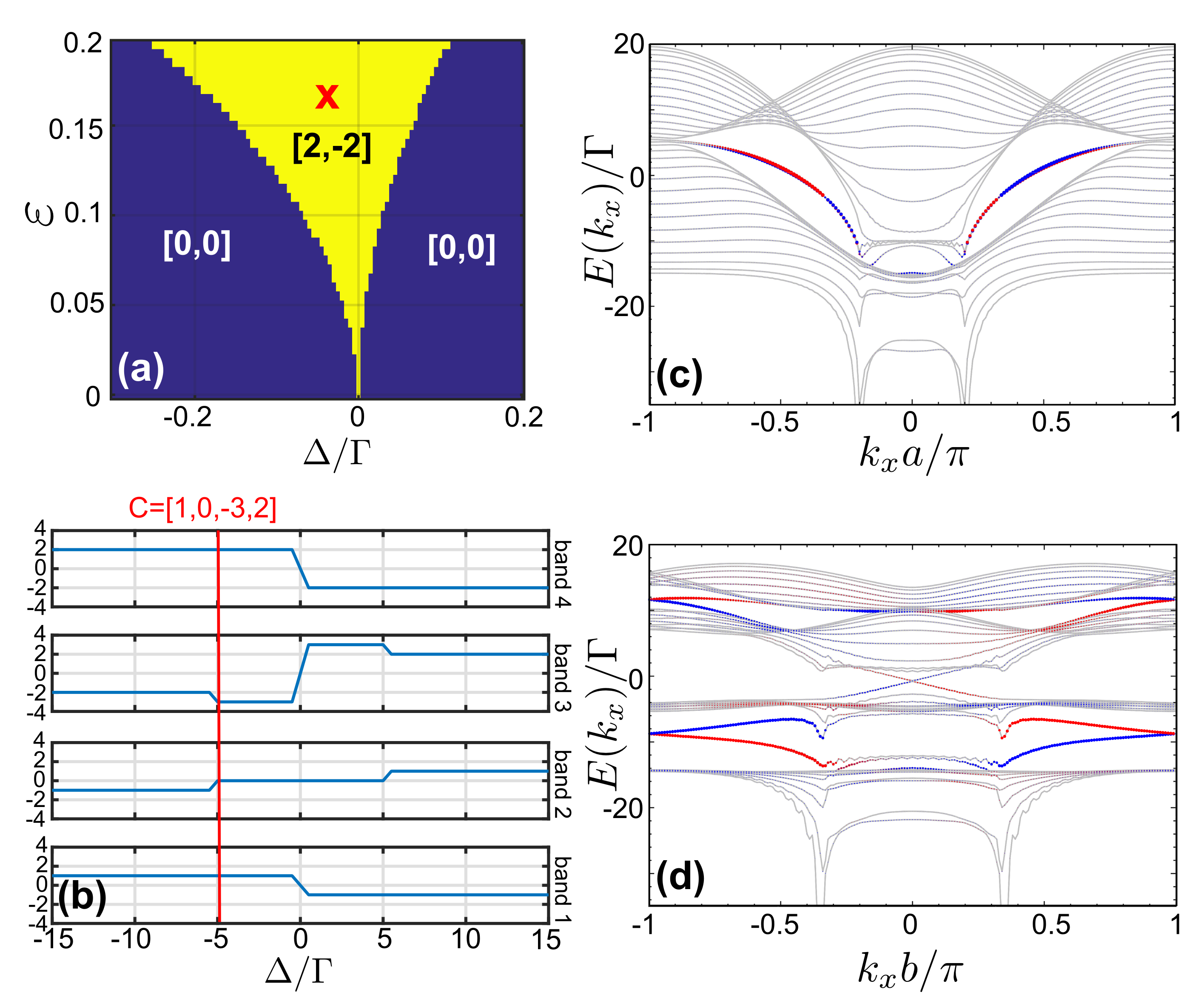}\\
 \caption{Phase diagram in the $\Delta-\varepsilon$ plane for the SL \textbf{(a)} and as a function of $\Delta$ for the HL {\bf (b)}. The numbers in the square brackets indicate the Chern numbers and are ordered from the lowest to highest energy bands.
 \textbf{(c)} and {\bf (d):} The spectrum in the infinite strip geometry for the SL ($N_y=15$ atoms, $\varepsilon=0.17$, $\Delta/\Gamma=-0.05$) and HL ($N_y=20$ atoms, $\Delta/\Gamma=-5$,  bearded zigzag boundary conditions \cite{Wang16}). Note that with these boundary conditions $b=\sqrt{3}a$ denotes the spacing between neighboring unit lattice sites in the $y$ direction. The corresponding points in the phase diagrams are denoted by the red cross (a) and red line (b), respectively. The size of the circles indicate the weight of the given eigenmode on one (red) or the other (blue) edge of the strip.}\label{fig:chern}
\end{figure}

The resulting phase diagram is shown in Fig. \ref{fig:chern}(a) (SL) and Fig. \ref{fig:chern}(b) (HL). We note that in the SL case it is required that $\varepsilon \neq 0$ in order to access the topologically non-trivial region and that in HL several topologically distinct regions can be accessed by tuning $\Delta$ alone (i.e., $\varepsilon=0$).

A hallmark of the non-trivial topology of the bulk is the appearance of edge states in finite-size systems. Considering an infinite strip keeping the system finite in the $y$ direction we plot the corresponding band structure in Fig. \ref{fig:chern}(c) (SL) and Fig. \ref{fig:chern}(d) (HL), where the edge states appear in the bulk gaps. Here, the size of the filled circles corresponds to the weight of the given eigenmode on one (red) or the other (blue) edge of the infinite strip.
	
We note that in the SL case the edge states are nearly degenerate and ''energetically hidden'' \cite{Weber15,Peter15} making it difficult to access them experimentally. In the HL case, relatively well separated bands form. Moreover, it is easy to verify that in the example shown indeed the Hall conductivity or, equivalently, the sum of the Chern numbers of the filled bands below a band gap, determines the net number of edge states crossing that gap \cite{Hatsugai93,Liu12}.

\noindent\textit{Finite system: quasi-momentum and driven-dissipative dynamics.} One important consequence of the long-ranged character of the interactions in the context of topological systems is that the bulk-edge correspondence, well established for systems with short-range potentials does not necessarily hold, as we exemplify in the following.

In Fourier space, the edge modes can be seen as modes spanning between different energy bands. In a finite system, to obtain a similar measure of whether any given mode has a chiral edge conductivity associated with it, we define a \textit{quasi-momentum} following a similar approach to \cite{Hafezi13,Weber15}: When an excitation hops from site to site, it accumulates a phase. If this phase of hopping is constant around the edge of the lattice, the hopping occurs around the edge in a given direction. If the hopping phase is random between different lattice sites, however, there is no preferred direction of hopping. We therefore define the \textit{quasi-momentum} $q$ as the average phase difference between neighbouring lattice sites around the edge of the lattice:
\begin{equation} \label{eq:quasimomentum}
q = \frac{1}{N_{\text{edge}}} \text{Im}\left[\text{ln} \sum_{l=1}^{N_{\text{edge}}}
\frac{\mathbf{d}_{l}^{\dag}\mathbf{d}_{l+1}}
{|\mathbf{d}_{l}^{\dag}\mathbf{d}_{l+1}|}\right],
\end{equation}
where $N_{\text{edge}}$ is the number of atoms in the outermost edge of the lattice. In Fig. 3(a) we show the quasi-momentum spectrum for a HL of an hexagonal shape with a total of 486 atoms \footnote{The effects we show in this work can also be observed in smaller lattices.}. We have overlayed the band structure results for an infinite strip with the same number of atoms in the $y$ direction for comparison.

\begin{figure}
\centering
 \includegraphics[width=1\columnwidth]{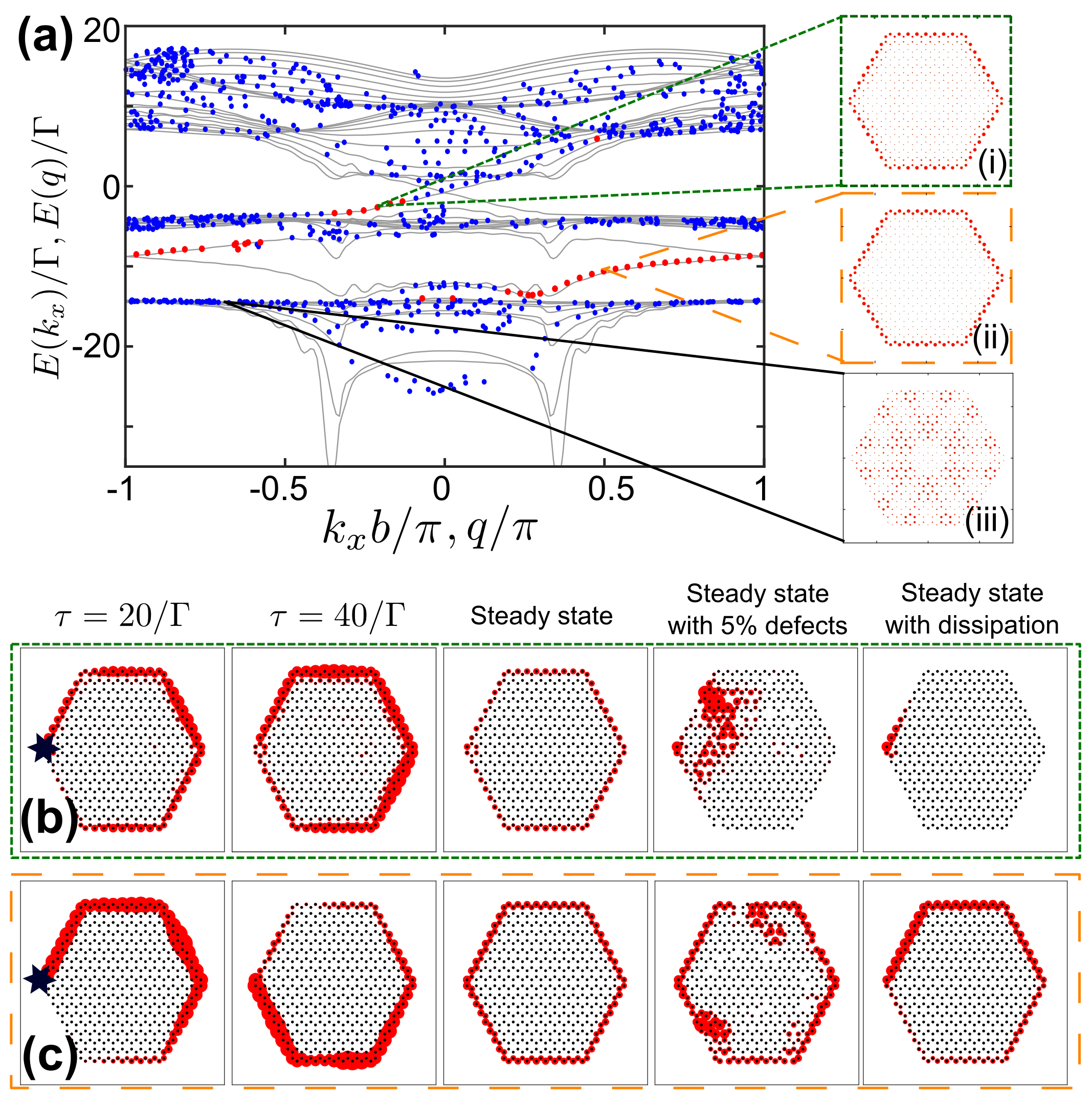}\\
 \caption{\textbf{(a):} Energy vs. quasi-momentum of a finite HL of hexagonal shape with the total of 486 sites (dots) together with the spectrum of the infinite strip geometry [grey lines, same parameters as in Fig. \ref{fig:chern}(d)]. The blue (red) dots represent states with less (more) than 95\% spatial support on the edge of the lattice (see text for details). {\bf(i-iii)} show the spatial configuration of the indicated eigenstates. {\bf (b)} and {\bf (c):} Excitation dynamics and the steady state (with defects and dissipation) under the driving of a single atom on the edge (indicated by the star in the leftmost panels) with detuning $-2\Gamma$ (b) and $-10\Gamma$ (c).}\label{fig:finite}
\end{figure}

Another defining characteristic of an edge state is that the vast majority of its weight is on the physical edge of the system. Writing an eigenstate of the system as $\ket{\varphi} = \sum_{\mu,j} c^j_{\mu} \ket{\mu}_j$, where $\mu \in \{-,+\}$ and $j$ index the internal degrees of freedom and the spatial position, respectively, we quantify the weight of the eigenstate on the edge of the system as $w = \sum_{\mu,j\,\in\, {\rm edge}} |c^j_{\mu}|^2$. The blue and red dots in Fig. \ref{fig:finite}(a) correspond to states where $w<0.95$ and $w>0.95$, respectively. The spatial configuration of two possible edge states from the middle and lowest bulk gap is shown in Figs. \ref{fig:finite}(i) and (ii) [which we denote as states (i) and (ii) in the following]. A bulk state is shown for comparison in Fig. \ref{fig:finite}(iii). Here, the size of the red circles correspond to the weight of the state at a given site.

In order to further characterize the edge states in the finite system, we model the dynamics of an excitation in the lattice under the Hamiltonian (\ref{eq:H}) and weak driving. Specifically, we drive a single atom on the leftmost corner of the HL [marked by a star in the leftmost panels of Figs. \ref{fig:finite}(b) and (c)] with weak driving strength and we choose the detuning to be $-2\Gamma$ and $-10\Gamma$ [Figs. \ref{fig:finite}(b) and (c), respectively], so that we resonantly address the states (i) and (ii), respectively. We observe that initially in both (b) and (c) the excitation dynamics is chiral with clockwise propagation before the steady state is reached.

Next, as one of the most appealing features of the edge states is their robustness against local perturbations, we simulate the excitation dynamics including $5\%$ of uniformly distributed lattice defects, i.e., empty sites due to an imperfect loading of atoms [fourth panels in Figs. \ref{fig:finite}(b,c)]. We observe a clear difference in the excitation propagation when driving the state (i), where the propagation is clearly interrupted, and the state (ii), where the excitations still populate a significant part of the edge. This strongly indicates that only the second one, state (ii), is an actual edge state. This is in contrast to what one might expect from the equilibrium bulk analysis, where the edge state conductivity, proportional to the sum of the Chern numbers of the filled lower lying bands, is the same for both states (i) and (ii) [see the red line in Fig. \ref{fig:chern}(b)].

Finally, as the mechanism that is responsible for the existence of coherent interactions is also responsible for the dissipation in the system, we simulate the dynamics above including the dissipative term (\ref{eq:diss}). Remarkably, we find that while in the case (i) the dissipation makes the excitation decay after only a few sites, when exciting the edge state (ii) the excitation becomes sub-radiant, i.e., it decays with a rate much smaller than the single atom decay rate $\Gamma$.

\noindent\textit{Conclusions and outlook.} We have found the existence of topologically non-trivial phases in a dense atomic two-dimensional lattice system coupled to the radiation field. We show that one can excite edge states that allow for the transport of an excitation over long distances along the edge of the lattice that are robust to the presence of defects. Moreover, these edge states are remarkably long-lived due to the collective character of the dissipation (a detailed analysis of this effect will be the aim of further investigations). Finally, we have found that, due to the long-range character of the interactions, the bulk-boundary relations, well established for topological insulators with short-range interactions, are not generally valid in our setting.

\begin{acknowledgments}
The authors thank M. Marcuzzi and S. Weber for useful discussions. The research leading to these results has received funding from the European Research Council under the European Union's Seventh Framework Programme (FP/2007-2013) / ERC Grant Agreement No. 335266 (ESCQUMA) and the EPSRC Grant No. EP/M014266/1. R.J. Bettles acknowledges funding from the UK EPSRC (Grant No. EP/L023024/1). B. Olmos was supported by the Royal Society and EPSRC grant DH130145. C.S. Adams also acknowledges support from FET-PROACT project ‘RySQ’ (H2020-FETPROACT-2014-640378-RYSQ).
\end{acknowledgments}

\appendix

\section{Coupling to the multimode radiation field}

The atomic system is coupled to the quantized multimode radiation field. Within the dipole approximation, the Hamiltonian that describes the dynamics of the atomic ensemble and the radiation field reads $H_\mathrm{af}=\sum_{j=1}^N\hbar\omega_\mathrm{a}\mathbf{b}_j^\dagger\cdot\mathbf{b}_j +\sum_{\mathbf{q}\lambda}\hbar\omega_{\mathbf{q}}a_{\mathbf{q}\lambda}^\dagger a_{\mathbf{q}\lambda}+i\hbar\sum_{j=1}^N\sum_{\mathbf{q}\lambda}
\mathbf{g}_{\mathbf{q}\lambda}\cdot\left(a_{\mathbf{q}\lambda}^\dagger
\mathbf{s}_j e^{-i\mathbf{q}\cdot\mathbf{r}_j}-
\mathbf{s}_ja_{\mathbf{q}\lambda}e^{i\mathbf{q}\cdot\mathbf{r}_j}\right)$. The first term represents the atomic energy, with $\hbar\omega_\mathrm{a}=2\pi\hbar c/\lambda$ being the energy difference between the ground ($\left|g\right>$) and the three degenerate $J=1$ manifold states ($\left|-\right>$, $\left|0\right>$ and $\left|+\right>$ for $m_J=-1,0$ and $+1$, respectively) and the vector transition operator for the $j$-th atom (located at $\mathbf{r}_j$) being defined as $\mathbf{b}_j=\left(\left|g\right>_j\!\left<+\right|,\left|g\right>_j\!\left<-\right|,\left|g\right>_j\!\left<0\right|\right)$. The second term of the Hamiltonian represents the radiation field, and here $\hbar\omega_{\mathbf{q}}$ is the energy of a photon with momentum $\mathbf{q}$ and polarization $\lambda$ and $a_{\mathbf{q}\lambda}$ is the annihilation operator of such a photon ($[a_{\mathbf{q}\lambda},a_{\mathbf{q}'\lambda'}^\dag] =\delta_{\mathbf{q}\mathbf{q}'}\delta_{\lambda\lambda'}$). The last term represents the coupling between the two systems, with $\mathbf{s}_j=\mathbf{b}_j^\dag+\mathbf{b}_j$ and the coupling coefficient being given by $\mathbf{g}_{\mathbf{q}\lambda}=d\sqrt{\frac{\omega_{\mathbf{q}}}{2\epsilon_0\hbar
V}} \hat{e}_{\mathbf{q}\lambda}$, where $V$ is the quantization volume, $d$ the modulus of the transition dipole moment between the ground and the three degenerate excited states, and $\hat{e}_{\mathbf{q}\lambda}$ the unit polarization vector of the photon ($\mathbf{q}\cdot\hat{e}_{\mathbf{q}\lambda}=0$).

Within the dipole and Born-Markov approximations, and tracing out the environment, one obtains the quantum master equation for the atomic density matrix $\rho$, $\dot{\rho}=-\frac{i}{\hbar}\left[H,\rho\right]+{\cal D}(\rho)$, which describes the dynamics of the atomic degrees of freedom \cite{Lehmberg70,Agarwal70,James93}. The many-body Hamiltonian that describes the coherent evolution of the system reads $H=\hbar\sum_{j\neq l}\mathbf{b}^\dag_{j}\cdot\overline{W}_{jl}\cdot\mathbf{b}_{l}$, characterized by the coefficient matrix
\begin{equation}\label{eq:V}
  \overline{W}_{jl}=\left(\begin{array}{ccc}
    A_{jl} & B_{jl}e^{-2i\phi_{jl}} & 0\\
    B_{jl}e^{2i\phi_{jl}} & A_{jl} & 0\\
    0 & 0 & C_{jl}
  \end{array}\right)
\end{equation}
with $A_{jl}$ and $B_{jl}$ being defined in the main text [Eqs. (5) and (6), respectively] and
\begin{eqnarray}
  C_{jl}&=&\frac{3\Gamma}{2}\left[-\frac{\sin{\kappa_{jl}}}{\kappa_{jl}^2} -\frac{\cos{\kappa_{jl}}}{\kappa_{jl}^3}\right]\label{eq:CC}
\end{eqnarray}
for $j\neq l$. The dissipator ${\cal D}(\rho)$ reads ${\cal D}(\rho) = \sum_{jl}\mathbf{b}_j\cdot\overline{\gamma}_{jl} \cdot\rho\mathbf{b}_l^\dag-\frac{1}{2}\left\{\mathbf{b}_j^\dag\cdot\overline{\gamma}_{jl} \cdot\mathbf{b}_l,\rho\right\}$, where
\begin{equation}\label{eq:diss1}
  \overline{\gamma}_{jl}=\left(\begin{array}{ccc}
    A'_{jl} & B'_{jl}e^{-2i\phi_{jl}} & 0\\
    B'_{jl}e^{2i\phi_{jl}} & A'_{jl} & 0\\
    0 & 0 & C'_{jl}
  \end{array}\right),
\end{equation}
with $A'_{jl}$ and $B'_{jl}$ being defined in the main text [Eqs. (7) and (8), respectively] and
\begin{eqnarray}
  C'_{jl}&=&3\Gamma\left[-\frac{\cos{\kappa_{jl}}}{\kappa_{jl}^2} +\frac{\sin{\kappa_{jl}}}{\kappa_{jl}^3}\right]\label{eq:CCp}
\end{eqnarray}
It is clear from the form of Eqs. (\ref{eq:V}) and (\ref{eq:diss1}) that the dynamics of the internal level $m_J=0$ is disconnected from the other two. We will consider thus only the dynamics on the subspace formed by the states $m_J=\pm1$.

\section{Ewald summation}

\subsection{General notions}
We briefly review the Ewald summation technique \cite{Bonsall77,Kantorovich04} introduced originally to cope with the evaluation of the electrostatic potential energy of a gas of electrons interacting through Coulomb force. Here we wish to evaluate the discrete sum (\ref{eq:Wk}). Considering two-dimensional lattices with unit cell defined by two lattice vectors $(\mathbf{a}_1, \mathbf{a}_2)$, a vector ${\mathbf r}$ connecting two atoms $\alpha$ and $\beta$ can be written as
\begin{equation}
	{\mathbf r} = {\pmb \tau}_{\alpha \beta} + {\bf T},
	\label{eq:r}
\end{equation}
where ${\bf T} = (m \mathbf{a}_1, n \mathbf{a}_2),\; m,n \in \mathbb{Z}$ is the lattice translation vector, ${\pmb \tau}_{\alpha \beta} = {\pmb \tau}_{\beta} - {\pmb \tau}_{\alpha}$ and ${\pmb \tau}_\alpha$ denotes the position of the $\alpha$-th atom in the unit cell. We also define the reciprocal lattice vectors in the usual way as ${\mathbf G} = (m \mathbf{b}_1, n \mathbf{b}_2),\; m,n \in \mathbb{Z}$, where $\mathbf{b}_j,\;j=1,2$ are the primitive reciprocal lattice vectors. 

Let us consider an algebraically decaying function of the atomic separation $v({\mathbf r})$ and say we want to evaluate the sum 
\begin{equation}
	V({\bf k}) = \sum_{{\mathbf r}\neq 0} {\rm e}^{i {\bf k r}} v({\bf r})
	\label{eq:Vk direct}
\end{equation}
The Ewald's trick consists of separating the contributions of $v$ to the sum to the short (S) and long (L) ranged part by introducing a regulating function $\xi$ through the identity
\begin{equation}
	1 = {}^S\xi(r) + {}^L\xi(r),
\end{equation}
so that
\begin{equation}
	V({\bf k}) = {}^S V({\bf k}) + {}^L V({\bf k}) - {}^{\rm self} V({\bf k}),
	\label{eq:Vk pot}
\end{equation}
where
\begin{subequations}
\begin{align}
	{}^S V({\bf k}) &= \sum_{\mathbf r\neq 0} {\rm e}^{i {\bf k r}} {}^S\xi(r) v({\bf r}) \label{eq:FS}\\
	{}^L V({\bf k}) &= \sum_{\mathbf r} {\rm e}^{i {\bf k r}} {}^L\xi(r) v({\bf r}) \label{eq:FL}\\	
	{}^{\rm self} V({\bf k}) &= \lim_{r \to 0} {\rm e}^{i {\bf k r}} {}^L\xi(r) v({\bf r}). \label{eq:Fself}
\end{align}
\end{subequations}
The function $\xi$ is chosen such that ${}^S\xi(r) v({\bf r})$ decays exponentially fast as $r \rightarrow \infty$ and (\ref{eq:Fself}) is finite. Due to the fast decay of ${}^S\xi$, the short-range part can be readily obtained by evaluating the sum in (\ref{eq:FS}) in real space. We are thus left with the evaluation of (\ref{eq:FL}). 

We proceed as follows. Using (\ref{eq:r}), (\ref{eq:FL}) becomes
\begin{widetext}
\begin{eqnarray}
	{}^L V({\bf k}) &=& \sum_{\alpha,\beta \in {\rm unit\;cell}} \sum_{\bf T} {\rm e}^{i {\bf k}({\pmb \tau}_{\alpha \beta} + {\bf T})} {}^L \xi(|{\pmb \tau}_{\alpha \beta} + {\bf T}|) v({\pmb \tau}_{\alpha \beta} + {\bf T}) \nonumber \\
	&=& \sum_{\alpha,\beta \in {\rm unit\;cell}} {\rm e}^{i {\bf k}{\pmb \tau}_{\alpha \beta}} \int {\rm d}^2{\mathfrak r} \sum_{\bf T} \delta^{2}({\boldsymbol{\mathfrak r}}-{\bf T}) {\rm e}^{i {\bf k}{\boldsymbol{\mathfrak r}} } {}^L \xi(|{\pmb \tau}_{\alpha \beta} + {\boldsymbol{\mathfrak r}}|) v({\pmb \tau}_{\alpha \beta} + {\boldsymbol{\mathfrak r}})  \nonumber \\
	&=& \sum_{\bf G} \sum_{\alpha,\beta \in {\rm unit\;cell}} {\rm e}^{i {\bf k}{\pmb \tau}_{\alpha \beta}} \int {\rm d}^2 {\mathfrak r} {\rm e}^{i {\bf k + G}{\boldsymbol{\mathfrak r}} } {}^L \xi(|{\pmb \tau}_{\alpha \beta} + {\boldsymbol{\mathfrak r}}|) v({\pmb \tau}_{\alpha \beta} + {\boldsymbol{\mathfrak r}})  \nonumber \\
	&=&  \sum_{\bf G} \sum_{\alpha,\beta \in {\rm unit\;cell}} {\rm e}^{-i {\bf G}{\pmb \tau}_{\alpha \beta}} \int {\rm d}^2 r {\rm e}^{i {\bf \tilde{G}}{\bf r}} {}^L \xi(r) v({\bf r}),
	\label{eq:FL transf}
\end{eqnarray}
\end{widetext}
where we in the third line we have used the identity (normalizing the volume of the unit cell to 1)
\begin{equation}
	\sum_{\bf T} \delta^{2}({\boldsymbol{\mathfrak r}}-{\bf T}) = \sum_{\bf G} {\rm e}^{i {\bf G}{\bf r}} 
\end{equation}
and in the last line we made a substitution of variables ${\bf r} = {\pmb \tau}_{\alpha \beta} + {\boldsymbol{\mathfrak r}}$ and defined $\widetilde{\bf G} = {\bf k + G}$.

Next, due to the form of (\ref{eq:W}),(\ref{eq:A}),(\ref{eq:B}), we consider the following functions $v$
\begin{subequations}
	\begin{align}
		{}_{1,n}v({\bf r}) &= \frac{\cos \kappa r}{\kappa r} {\rm e}^{i n \phi({\bf r})} \label{eq:1nv}\\
		{}_{2,n}v({\bf r}) &= \frac{\sin \kappa r}{(\kappa r)^2} {\rm e}^{i n \phi({\bf r})} \label{eq:2nv}\\
		{}_{3,n}v({\bf r}) &= \frac{\cos \kappa r}{(\kappa r)^3} {\rm e}^{i n \phi({\bf r})}, \label{eq:3nv}
	\end{align}	
\end{subequations}
where $n=0,2$ and $\kappa = 2\pi/\lambda$. 

\subsection{Analytical approach to the $1/r$ term: the divergence}

\begin{center}
	\begin{figure}[h!]	
  	\includegraphics[width=8cm]{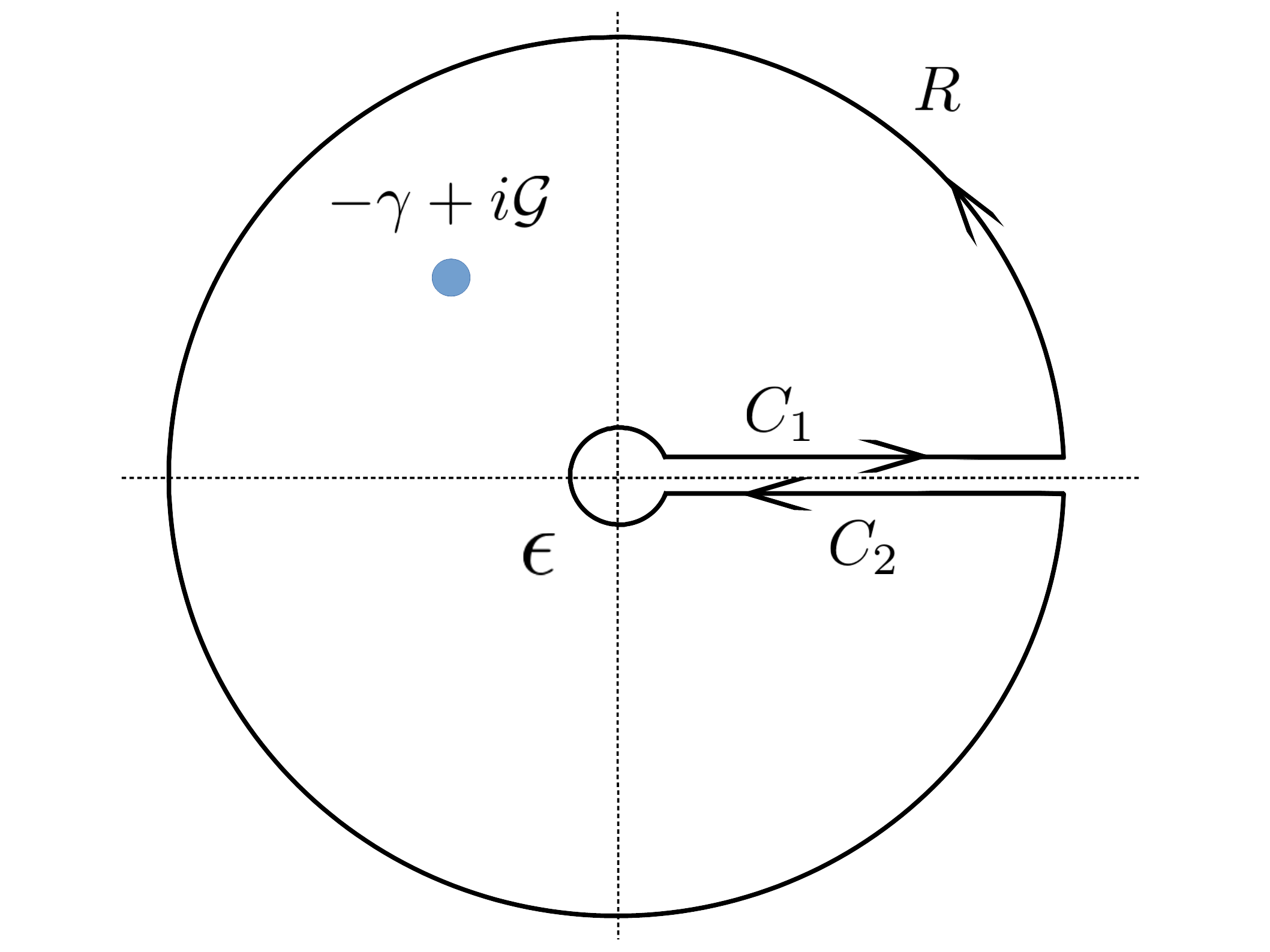} 
  	\caption{integration contour}
  	\label{fig:contour}
 	\end{figure}
\end{center}

We start with the evaluation of the integral in (\ref{eq:FL transf}) with the function ${}_{1,n}v({\bf r})$
\begin{eqnarray}
	{}_{1,n}\mathfrak{V} &\equiv & \int {\rm d}^2 r {\rm e}^{i {\bf \widetilde{G}}{\bf r}} {}^L \xi(r) {}_{1,n}v({\bf r}) \nonumber \\
	&=& \kappa^{-1} \int_0^\infty {\rm d}r \left(1-{\rm e}^{-\gamma r} \right) \cos(\kappa r)\!\!\! \int_0^{2\pi}\!\!\! {\rm d} \varphi {\rm e}^{i ({\widetilde{G}}r \cos{(\varphi-\eta)} + n \phi({\bf r})) } \nonumber \\
	&=& \kappa^{-1} \int_0^\infty {\rm d}r \left(1-{\rm e}^{-\gamma r} \right) \cos(\kappa r) J_n(\widetilde{G}r),
\end{eqnarray}
where we have chosen ${}^L\xi(r) = 1-{\rm e}^{-\gamma r}$, $\gamma >0$ is an arbitrary control parameter, $\eta$ is the polar angle of $\widetilde{\bf G}$ and $J_n$ is the $n$-th Bessel function of the first kind. Next, we split the integration as
\begin{equation}
	{}_{1,n}\mathfrak{V} ={}_{1,n}\mathfrak{V}_0^c+{}_{1,n}\mathfrak{V}_c^\infty,
\end{equation}
where
\begin{equation}
	{}_{1,n}\mathfrak{V}_a^b=\kappa^{-1} \int_a^b {\rm d}r \left(1-{\rm e}^{-\gamma r} \right) \cos(\kappa r) J_n(\widetilde{G}r).
\end{equation}
Finally, we substitute the asymptotic expression for the Bessel function (for large $\widetilde{G}r$)
\beq
	J_n(\widetilde{G}r) \approx \sqrt{\frac{2}{\pi \widetilde{G}}} \frac{1}{\sqrt{r}} \cos \left( \widetilde{G} r -\phi_n\right),
	\label{eq:Bessel approx}
\eeq
where $\phi_n = \frac{\pi}{4}(2 n + 1)$, to ${}_{1,n}\mathfrak{V}_c^\infty$ so that
\beq
	{}_{1,n}\mathfrak{V}_c^\infty \approx I_0^\infty - I_0^c,
\eeq
where
\beq
	I_a^b = \frac{1}{\kappa}\sqrt{\frac{2}{\pi \widetilde{G}}} \int_c^d\!\! {\rm d}r \frac{1-{\rm e}^{-\gamma r}}{\sqrt{r}} \cos(\kappa r) \cos(\widetilde{G}r-\phi_n).
	\label{eq:Iab}
\eeq
Importantly, we note, that only the term $I_0^\infty$ can give rise to the divergence in (\ref{eq:Vk pot}), which we now analyze in detail. Substituting the identity
\beq
	\frac{1}{\sqrt{r}} = \frac{1}{\sqrt{\pi}} \int_0^\infty {\rm d}t t^{-\frac{1}{2}} {\rm e}^{-t r}
\eeq
and using
\beq
	\cos{a}\cos{b} = \frac{1}{2} \left[ \cos(a+b) + \cos(a-b) \right]
\eeq
we get
\begin{widetext}
\beq
	\kappa \pi \sqrt{\frac{\widetilde{G}}{2}} I_0^\infty = \int_0^\infty {\rm d}t t^{-\frac{1}{2}} \int_0^\infty {\rm d}r \left( 1-{\rm e}^{-\gamma r} \right) {\rm e}^{-t r} \left[ \cos((\widetilde{G} + \kappa)r - \phi_n) + \cos((\widetilde{G} - \kappa)r - \phi_n) \right].
	\label{eq:I 0 infty}
\eeq
\end{widetext}
This amounts to evaluation of the integral
\beqa
	II &=& \int_0^\infty {\rm d}t t^{-\frac{1}{2}} \int_0^\infty {\rm d}r \left( 1-{\rm e}^{-\gamma r} \right) {\rm e}^{-t r} \cos(\mathcal{G} r - \phi_n) \nonumber \\
	&=& \frac{{\rm e}^{-i \phi_n}}{2} \int_0^\infty {\rm d}t \frac{1}{\sqrt{t}} \int_0^\infty {\rm d}r  \left[ {\rm e}^{-t r} - {\rm e}^{-(t+\gamma) r} \right] {\rm e}^{i \mathcal{G} r} + {\rm c.c.} \nonumber \\
	&=& \frac{{\rm e}^{-i \phi_n}}{2} \int_0^\infty {\rm d}t \frac{1}{\sqrt{t}} \left( \frac{1}{t-i \mathcal{G}} - \frac{1}{t+\gamma-i \mathcal{G}}\right) + {\rm c.c.},
	\label{eq:II}
\eeqa
where $\mathcal{G} = \widetilde{G} \pm \kappa$. The final integration in (\ref{eq:II}) can be done by methods of contour integration along the contour depicted in Fig. \ref{fig:contour}, i.e.
\beq
	\oint_C = \int_{C_1} + \int_R + \int_{C_2} + \int_\epsilon.
	\label{eq:cont int}
\eeq
Using the property of the branch cut, the integration along $C_2$ becomes
\beqa
	\int_{C_2} &=& \int_R^\epsilon {\rm d}z \frac{1}{\sqrt{z}} \left( \frac{1}{z-z_0} - \frac{1}{z-z_1} \right) \nonumber \\
	&=& \int_R^\epsilon {\rm d}z \frac{1}{{\rm e}^{\frac{1}{2}(\log |z| + i 2 \pi)}} \left( \frac{1}{z-z_0} - \frac{1}{z-z_1} \right) \nonumber \\
	&=& \int_R^\epsilon {\rm d}z \frac{1}{-\sqrt{z}} \left( \frac{1}{z-z_0} - \frac{1}{z-z_1} \right) \nonumber \\
	&=& \int_\epsilon^R {\rm d}z \frac{1}{\sqrt{z}} \left( \frac{1}{z-z_0} - \frac{1}{z-z_1} \right).
\eeqa
Omitting the c.c. term, we express the leftmost part of (\ref{eq:II}) as
\beqa
	II_- &=& \frac{{\rm e}^{-i \phi_n}}{2} \int_0^\infty\!\!\! {\rm d}z \frac{1}{\sqrt{z}} \left( \frac{1}{z-i \mathcal{G}} - \frac{1}{z+\gamma-i \mathcal{G}}\right)\\
	 &=& \int_0^\infty {\rm d}z g(z) = \int_0^\infty {\rm d}z \frac{f(z)}{z-z_0} - \frac{f(z)}{z-z_1}.
\eeqa
It can be immediately seen, that $\int_R \rightarrow 0$ since $g(z) \rightarrow R^{-\frac{3}{2}}$ in the limit of large $|z|$. Next, since $\gamma>0$, assuming $\mathcal{G} \neq 0$ and substituting $z=\epsilon {\rm e}^{i \varphi}$, we get from the estimation lemma
\beq
	\left| \int_\epsilon \right| \leq |g(z)| 2\pi \epsilon \propto 2 \pi \frac{\epsilon}{\sqrt{\epsilon}} \xrightarrow[]{\epsilon \rightarrow 0^+} 0.
\eeq
We are thus left with
\beqa
	II_- &=& \frac{1}{2} \oint_C = i \pi \left( f(z_0) - f(z_1) \right)\label{eq:II minus}\\\nonumber
	 &=& i \pi \frac{{\rm e}^{-i \phi_n}}{2} \left[ \frac{1}{\sqrt{i \mathcal{G}}} - \frac{1}{\sqrt{-\gamma + i \mathcal{G}}} \right],
\eeqa
where we have defined the function $f$ and the poles
\beqa
	f(z) &=& \frac{{\rm e}^{-i \phi_n}}{2}\frac{1}{\sqrt{z}} \nonumber \\
	z_0 &=& i \mathcal{G} \nonumber \\
	z_1 &=& -\gamma + i \mathcal{G}.	
\eeqa
\begin{widetext}
Plugging (\ref{eq:II minus}) back to (\ref{eq:II}) and evaluating the complex conjugated term we get
\beqa
	\frac{2}{\pi} II &=& i {\rm e}^{-i \phi_n} \left[ \frac{1}{\sqrt{i \mathcal{G}}} - \frac{1}{\sqrt{-\gamma + i \mathcal{G}}} \right] + {\rm c.c} \nonumber \\
	&=& i \cos(\phi_n) \left[ \frac{1}{\sqrt{i \mathcal{G}}} - \frac{1}{\sqrt{-i \mathcal{G}}} + \frac{1}{\sqrt{-\gamma-i \mathcal{G}}} - \frac{1}{\sqrt{-\gamma + i \mathcal{G}}} \right]
+ \sin(\phi_n) \left[ \frac{1}{\sqrt{i \mathcal{G}}} + \frac{1}{\sqrt{-i \mathcal{G}}} - \frac{1}{\sqrt{-\gamma-i \mathcal{G}}} + \frac{1}{\sqrt{-\gamma + i \mathcal{G}}} \right] \nonumber \\
		&=& 2 \left\{ \frac{\cos(\phi_n) \sin(\frac{1}{2} {\rm arg}(i \mathcal{G}))}{|\mathcal{G}|^\frac{1}{2}} - \frac{\cos(\phi_n) \sin(\frac{1}{2} {\rm arg}(\gamma + i \mathcal{G}))}{|\gamma + i \mathcal{G}|^\frac{1}{2}}  + \frac{\sin(\phi_n) \cos(\frac{1}{2} {\rm arg}(i \mathcal{G}))}{|\mathcal{G}|^\frac{1}{2}} + \frac{\sin(\phi_n) \cos(\frac{1}{2} {\rm arg}(\gamma + i \mathcal{G}))}{|\gamma + i \mathcal{G}|^\frac{1}{2}} \right\}.
		\label{eq:II expansion}
\eeqa

Next, we note that this result can be further simplified in specific cases: for $\mathcal{G} \in \mathbb{R}$ and $n=0,2$ respectively, we get the following options
\beqa
	&n=0:&\; \phi_0 = \frac{\pi}{4} \rightarrow \sin(\frac{\pi}{4}) = \frac{1}{\sqrt{2}}; \;\;\; \cos(\frac{\pi}{4}) = \frac{1}{\sqrt{2}} \nonumber \\
	&n=2:&\; \phi_2 = \frac{5 \pi}{4} \rightarrow \sin(\frac{5 \pi}{4}) = \frac{1}{\sqrt{2}}; \;\;\; \cos(\frac{5 \pi}{4}) = -\frac{1}{\sqrt{2}} \nonumber \\
	&\mathcal{G}>0:&\; {\rm arg}(i \mathcal{G}) = \frac{\pi}{2} \rightarrow \sin(\frac{1}{2} {\rm arg}(i \mathcal{G})) = \frac{1}{\sqrt{2}}; \;\;\; \cos(\frac{1}{2} {\rm arg}(i \mathcal{G})) = \frac{1}{\sqrt{2}}  \nonumber \\
	&\mathcal{G}<0:&\; {\rm arg}(i \mathcal{G}) = -\frac{\pi}{2} \rightarrow \sin(\frac{1}{2} {\rm arg}(i \mathcal{G})) = -\frac{1}{\sqrt{2}}; \;\;\; \cos(\frac{1}{2} {\rm arg}(i \mathcal{G})) = \frac{1}{\sqrt{2}}.
	\label{eq:na}
\eeqa
Lets first consider the $n=0$ case. Using (\ref{eq:na}), the relation (\ref{eq:II expansion}) simplifies to
\beqa
	&\mathcal{G}>0:& \; \frac{1}{\pi} II_{n=0}(\mathcal{G}) = \frac{1}{|\mathcal{G}|^\frac{1}{2}} - \frac{1}{\sqrt{2}|\gamma + i \mathcal{G}|^\frac{1}{2}} \left( \sin(\frac{1}{2}{\rm arg}(\gamma + i \mathcal{G})) - \cos(\frac{1}{2}{\rm arg}(\gamma + i \mathcal{G})) \right) \nonumber \\
		&\mathcal{G}<0:& \; \frac{1}{\pi} II_{n=0}(\mathcal{G}) = - \frac{1}{\sqrt{2}|\gamma + i \mathcal{G}|^\frac{1}{2}} \left( \sin(\frac{1}{2}{\rm arg}(\gamma + i \mathcal{G})) - \cos(\frac{1}{2}{\rm arg}(\gamma + i \mathcal{G})) \right).
\eeqa
and similar expression is obtained for $n=2$. Those results can be combined to a single expression
\beq
	\frac{1}{\pi} II_n(\mathcal{G}) = \frac{(-1)^{n}}{|\mathcal{G}|^\frac{1}{2}} \theta(\mathcal{G}) - \frac{1}{\sqrt{2}|\gamma + i \mathcal{G}|^\frac{1}{2}} \left( \sin(\frac{1}{2}{\rm arg}(\gamma + i \mathcal{G})) - \cos(\frac{1}{2}{\rm arg}(\gamma + i \mathcal{G})) \right),
	\label{eq:II n=0}
\eeq
where $\theta$ is the Heaviside step function. Remarkably, the function (\ref{eq:II n=0}) is finite in the limit $\mathcal{G} \rightarrow 0^-$, but diverges as $\mathcal{G} \rightarrow 0^+$, which constitutes the main analytic result of this section.
\end{widetext}

\emph{Remark:} The discontinuity and one-sided divergence could be seen already in (\ref{eq:II}) by setting $\mathcal{G}=0$. This would result in the diverging integral in the vicinity of the origin, namely the part $\int_\epsilon \propto \int {\rm d}t \frac{1}{t\sqrt{t}}$ in (\ref{eq:cont int}), which, according to the estimation lemma, scales as $\propto \frac{1}{\sqrt{\epsilon}}$.

\subsection{Numerical implementation}

In the previous section we have found that (\ref{eq:FS}),(\ref{eq:Fself}) can be evaluated numerically, while (\ref{eq:FL}) can be written as
\begin{equation}
	{}^{\phantom{1,}L}_{1,n}V({\bf k}) = \sum_{\bf G} \sum_{\alpha,\beta \in {\rm unit\;cell}} {\rm e}^{-i {\bf G}{\pmb \tau}_{\alpha \beta}} \left[ {}_{1,n}\mathfrak{V}_0^c - I_0^c + I_0^\infty \right].
	\label{eq:1nV}
\end{equation}
Here, the expressions ${}_{1,n}\mathfrak{V}_0^c$ and $I_0^c$ are finite and can by obtained numerically. In principle similar decomposition can be obtained when considering the functions (\ref{eq:2nv}),(\ref{eq:3nv}), however we found it more convenient to evaluate the corresponding element (\ref{eq:Vk direct}) by direct summation in real space. 

The final subtlety is related to the fact, that for ${}^L \xi = 1-{\rm e}^{-\gamma r}$ the self-interaction term (\ref{eq:Fself}) is not defined for ${}_{1,n=2}v$ as it depends on the direction in which the origin is approached (which is due to the dependence of ${}_{1,n=2}v$ on the polar angle of $\bf r$). This problem can be circumvented by considering ${}^L \xi = 1-{\rm e}^{-\gamma r^2}$ instead in which case the self-interaction term vanishes. Here, it is still possible to evaluate the integral in (\ref{eq:Iab}) symbolically, which we then use in (\ref{eq:1nV}).

In (\ref{eq:1nV}) the tuning parameters $\gamma, c$ are chosen such that the approximations used (eq. (\ref{eq:Bessel approx})) are well satisfied.

%

\end{document}